\RequirePackage{lineno}
\documentclass[preprint,12pt,3p,twocolumn]{elsarticle}
\usepackage{graphicx}
\usepackage{epsfig}
\usepackage{dcolumn}
\usepackage{bm}
\usepackage{amssymb}
\usepackage{multirow}
\usepackage{subfigure}
\usepackage{atlasphysics}

\usepackage{preprintcover}
\PreprintCoverPaperTitle{Search for a heavy narrow resonance decaying to $e \mu$, $e \tau$, or $\mu \tau$
with the ATLAS detector in $\sqrt{s}= 7$~TeV $pp$ collisions at the LHC}
\PreprintIdNumber{CERN-PH-EP-2012-326}
\PreprintCoverAbstract{This letter presents the results of a search for a heavy particle decaying
into an $e^\pm \mu^\mp$, $e^\pm \tau^\mp$, or $\mu^\pm \tau^\mp$
final state in $pp$ collisions at $\sqrt{s}=7$ TeV.
The data were recorded with the ATLAS detector at the LHC during 2011 and correspond
to an integrated luminosity of 4.6 fb$^{-1}$.
No significant excess above the Standard Model expectation is observed, and
exclusions at 95\% confidence level are placed on the cross section times branching
ratio for the production of an $R$-parity-violating supersymmetric tau sneutrino.
These results considerably extend constraints from Tevatron
experiments.}
\PreprintJournalName{Physics Letters B}

\newcommand{\vmet}{\ensuremath{\vec{\slash\kern-.7emE}_{T}}}
\newcommand{\snutau}{\ensuremath{\tilde{\nu}_{\tau}}}

\newcommand{\zp}{\ensuremath{Z'}}
\newcommand{\msnt}{\ensuremath{m_{\tilde{\nu}_{\tau}}}}

\journal{Physics Letters B}

\begin{document}
\begin{frontmatter}

\title{Search for a heavy narrow resonance decaying to $e\mu$, $e\tau$, or $\mu\tau$ with the ATLAS detector
in $\sqrt{s} = 7$ TeV $pp$ collisions at the LHC}
\author{The ATLAS Collaboration}

\begin{abstract}
This letter presents the results of a search for a heavy particle decaying
into an $e^\pm \mu^\mp$, $e^\pm \tau^\mp$, or $\mu^\pm \tau^\mp$
final state in $pp$ collisions at $\sqrt{s}=7$ TeV.
The data were recorded with the ATLAS detector at the LHC during 2011 and correspond
to an integrated luminosity of 4.6 fb$^{-1}$.
No significant excess above the Standard Model expectation is observed, and
exclusions at 95\% confidence level are placed on the cross section times branching
ratio for the production of an $R$-parity-violating supersymmetric tau sneutrino.
These results considerably extend constraints from Tevatron
experiments.
\end{abstract}

\end{frontmatter}

\section{Introduction}
Neutrino oscillations show that lepton-flavour quantum numbers are not conserved
in Nature.
On the other hand, lepton-flavour violation (LFV) has
not been observed in the charged lepton sector,
where neutrino-induced LFV is predicted to be extremely small in the Standard Model (SM).
The study of possible LFV processes involving charged leptons is
an important topic in the search for physics beyond the SM.
One possible signature is the production of a particle
that decays to a pair of different flavour, opposite-sign leptons $e^\pm \mu^\mp$ ($e \mu$),
$e^\pm \tau^\mp$ ($e \tau$), or $\mu^\pm \tau^\mp$ ($\mu \tau$) (referred to generically
as $\ell \ell^{\prime}$).  
Since leptons with large transverse momenta are identified cleanly,
efficiently, and with good resolution, the ATLAS detector is well suited to a search
for this signature.
Many new physics models allow LFV in charged lepton interactions. 
For example, in $R$-parity-violating~(RPV) models of supersymmetry (SUSY)~\cite{rpv_sneutrino},
a sneutrino can have LFV decays to $\ell \ell^{\prime}$.
Models with additional gauge
symmetry can accommodate an $\ell \ell^{\prime}$ signature through LFV
decays of an extra gauge boson $\zp$~\cite{zprime_pap}.
This signature is also produced in the SM framework, for example,
$t\bar{t}$, $WW$, or $Z/\gamma^* \to \tau^- \tau^+$ production where the final-state
particles decay to leptons of different flavour.  These processes
typically have small cross sections,
and the $\ell \ell^{\prime}$ invariant mass~($m_{\ell \ell^{\prime}}$) lies predominantly
below the range favoured for new physics signals.

This letter reports on a search for a heavy particle decaying into the $e\mu$, $e \tau_\mathrm{had}$, or $\mu \tau_\mathrm{had}$
final state, where $\tau_\mathrm{had}$ is a $\tau$ lepton that decays hadronically.
The search uses 4.6~fb$^{-1}$ of 7~TeV $pp$ collision data taken with the \mbox{ATLAS} detector during 2011.
The results are interpreted in terms of the production via $d\bar{d}$ annhiliation and subsequent decay
of a tau sneutrino \snutau\ in RPV SUSY ($d\bar{d} \to \snutau \to \ell\ell^{\prime}$). 
Both the CDF and D0 Collaborations at the Tevatron collider have reported searches
for the RPV production and decay of a \snutau\ in the $e\mu$ channel~\cite{RPVTevatron}.
The CDF Collaboration also set limits in the $e \tau$ and $\mu \tau$ channels~\cite{RPVTevatron}.
This letter supersedes previous \mbox{ATLAS} searches for a high-mass resonance decaying to $e \mu$
based on 1~fb$^{-1}$ of 2011 data~\cite{ATLAS_emu} and extends
the search to $e \tau_\mathrm{had}$ and $\mu \tau_\mathrm{had}$ final states.

Precision low-energy searches, such as $\mu$ to $e$ conversion on nuclei,
rare muon decays, and rare tau decays, place limits on RPV couplings~\cite{low_energy}.  These limits often
depend on masses of supersymmetric particles that occur in loop diagrams and often need
to assume the dominance of certain couplings or pairs of couplings to extract limits.

\section{ATLAS detector}
The \mbox{ATLAS} experiment at the LHC employs a multipurpose particle
physics detector~\cite{atlas_detector} with a forward-backward symmetric
cylindrical geometry and near $4\pi$ coverage in
solid angle.\footnote{ATLAS uses a right-handed coordinate system with its origin
at the nominal interaction point (IP) in the centre of the detector
and the $z$-axis along the beam pipe. The $x$-axis points from the IP
to the centre of the LHC ring, and the $y$-axis points upward.
Cylindrical coordinates $(r,\phi)$ are used in the transverse (x,y) plane,
$\phi$ being the azimuthal angle around the beam pipe.
The pseudorapidity is defined in terms of the polar angle $\theta$ as $\eta=-\ln\tan(\theta/2)$.}
The inner tracking detector covers the pseudorapidity region
$|\eta|$~$<$~2.5 and consists of a silicon
pixel detector, a silicon microstrip detector, and
a transition radiation tracker.
The inner tracking detector is surrounded by a thin superconducting solenoid that provides a
2~T magnetic field and by a finely-segmented calorimeter with nearly full solid-angle coverage. 
The latter covers the pseudorapidity range $|\eta|$~$<$~4.9 and provides three-dimensional reconstruction 
of particle showers.
The electromagnetic compartment uses lead absorbers with liquid-argon as the active material.
This is followed by a hadronic compartment,
which uses scintillating tiles with iron absorbers in the central region and 
liquid-argon sampling with copper or tungsten absorbers for $|\eta|~>~1.7$. 
The muon spectrometer surrounds
the calorimeters and consists of three large superconducting toroids (each with eight coils),
a system of precision tracking chambers ($|\eta|$ $<$ 2.7), and detectors for triggering.

\section{Data and event selection}

\indent The data used in this analysis were recorded in 2011 
at a centre-of-mass energy of 7~TeV. Only data with stable run conditions
and operational tracking, calorimetry, and muon subdetectors are used.
This results in a data sample with an integrated luminosity of
4.6 fb$^{-1}$ with an estimated uncertainty of 3.9\%~\cite{lumi}. 
Events are required to satisfy a single-electron trigger for the $e\mu$ and $e \tau_\mathrm{had}$ searches
and a single-muon trigger for the $\mu \tau_\mathrm{had}$ search.
The nominal transverse momentum (\pt{}) threshold for the electron trigger was 20 or 22~GeV,
depending on the instantaneous luminosity, and was 18~GeV for the muon trigger.

An electron candidate is required to have
$\pt>25$~GeV and to lie in the pseudorapidity region $|\eta| < 2.47$,
excluding the transition region ($1.37 < |\eta| < 1.52$) between the barrel and endcap
calorimeters.
The \pt\ of the electron is calculated from the calorimeter energy and the direction
of the inner detector track.
A set of electron identification criteria based on the calorimeter shower shape, track quality,
transition radiation,
and track matching with the calorimeter energy deposition, referred to as `tight'~\cite{electronMedium},
is applied.
Two lepton isolation criteria are used to reduce backgrounds from hadronic jets. 
The calorimetric isolation criterion requires that
the transverse energy deposited within a cone of radius $\Delta R=\sqrt{(\Delta \eta)^2 + \Delta (\phi)^2}$~$=$~0.3
around the electron cluster, excluding the core energy deposited by the electron,
is less than 0.14 times the \pt\ of the candidate.
The tracking isolation criterion requires
the sum of the transverse momenta of tracks with
$\pt>1$~GeV within a cone of radius $\Delta R<0.3$ around the electron track,
excluding the electron track, is less than 0.13 times the \pt\ of the candidate.

A muon candidate must have reconstructed tracks in both the inner detector and the muon spectrometer.
The muon track parameters are a statistical combination of those for the inner detector and muon spectrometer
tracks.  The inner detector track is
required to have a pattern of hits consistent with a quality track.
Furthermore, the muon candidate must have 
$\pt>25$~GeV and be isolated, using similar criteria as for electrons: 0.14 times \pt\
for calorimetric isolation and 0.15 times \pt\ for tracking isolation.

Jets are reconstructed using the anti-$k_t$ jet clustering algorithm~\cite{antiKtTopo}
with a radius parameter of 0.4. 
Only jets with \pt~$>$~20~GeV and $|\eta|$~$<$~2.5 are considered. 
Leptons are retained only if they are separated from all 
jets by $\Delta R>0.4$.

Tau leptons are reconstructed through their hadronic decays.  The tau reconstruction
is seeded by anti-$k_t$ jets~\cite{antiKtTopo} with cone size $\Delta R  = 0.4$ and  jet $\pt > 10$~GeV.
Corrections depending on \pt\ and $\eta$ are applied
to the tau energy.  A boosted decision tree discriminator~\cite{tauRecon} efficiently selects
taus while rejecting backgrounds.
The variables used in the discriminator are $\Delta R$ between the tracks and the tau candidate, the impact
parameter significance of the tracks, the fraction of the \pt\ of the tau candidate carried by the tracks,
the number of tracks in an isolation annulus of 0.2~$<$~$\Delta R$~$<$~0.4, the width of the energy deposition
in the calorimeter,
energy isolation for cones of $\Delta R = 0.1$ and $\Delta R = 0.4$, and the invariant mass associated with the energy deposition. 
For this analysis, 
`medium' selection criteria as described in reference~\cite{tauRecon} are used.
This selection is about 60\% efficient at retaining taus that decay hadronically, as measured
in $Z\to \tau \tau$ decays, while
accepting 1 of 20 to 1 of 50 ordinary hadronic jets misidentified as tau candidates.
The reconstruction
efficiency for hadronic tau decays with three tracks drops significantly at large transverse momentum
as the tracks become more collimated.
This analysis therefore uses only tau candidates
with one track, which comprise 85\% of tau decays.
Tau candidates must have $\et$~$>$~20~GeV and
pseudorapidity in the range 0.03~$<$~$|\eta|$~$<$~2.5.  The lower limit excludes a
region where there is reduced coverage from the inner detector and calorimeters, which
greatly increases misidentification of electrons as hadronic tau decays.
To retain only taus that decay hadronically ($\tau_\mathrm{had}$),
candidates consistent with being
an electron or a muon are rejected.

The missing transverse energy ($\met$) is calculated from the vector sum of the transverse momenta
of all high-\pt\ objects (electrons, muons, photons, taus, and jets)
and all calorimeter energy clusters with $|\eta|$~$<$~4.5 not associated with those objects~\cite{missingET}.

Events are required to have exactly two lepton candidates with opposite sign and different
flavour, that is, $e \mu$, $e \tau_\mathrm{had}$, or $\mu \tau_\mathrm{had}$.  In addition,
each event must have at least one primary vertex with at least four tracks with $\pt>400$~MeV.
The two leptons are chosen to be back-to-back in $\phi$ by requiring that
the azimuthal angle between them satisfies $\Delta \phi_{\ell \ell^{\prime}}$~$>$~2.7.
Finally, for the $e \tau_\mathrm{had}$ and $\mu \tau_\mathrm{had}$ events, the $\pt$ of the electron or muon is required
to be greater than the $\et$ of the tau.

For $e \tau_\mathrm{had}$ and $\mu \tau_\mathrm{had}$ signal events, the presence of only one tau and the requirement of large
momentum relative to the tau mass implies that the neutrino from the tau decay
should point in nearly the same direction as the tau momentum and that
there are no other significant sources of $\met$.
The transverse components of the neutrino momentum are set equal to 
the components of the \met\ vector and the polar angle of the
neutrino momentum is set equal to the polar angle of the tau candidate's momentum.
The momentum of the tau candidate is corrected for the momentum of the neutrino
in the calculation of the $e \tau_\mathrm{had}$ and $\mu \tau_\mathrm{had}$ invariant mass.
This significantly reduces the width of the invariant mass distribution
for $e \tau_\mathrm{had}$ and $\mu \tau_\mathrm{had}$ pairs and
improves the search sensitivity, while making no significant changes to the background shapes.
For dilepton masses from 400~GeV to 2000~GeV,
the mass resolutions range from 2.5\% to 7.5\%, 2.2\% to 4.3\%, and 6.3\% to 9.0\% for
the $e\mu$, $e\tau_\mathrm{had}$, and $\mu \tau_\mathrm{had}$ decay modes, respectively.
The mass resolutions are dominated by the resolution of the transverse momenta of the leptons.
At high $\pt$, the transverse momentum resolution is best for electrons, whose \pt\ measurement
is based primarily on energy deposited in the electromagnetic calorimeter.
It is next best for taus, whose \pt\ measurement is based on electromagnetic and hadronic calorimeter
energy depositions.  It is the worst for muons, whose \pt\ measurement is from tracking.
The $\et$ of the tau used in the selection
requirements above is not corrected for the neutrino momentum.

\section{Backgrounds}
The SM processes that can produce an $\ell \ell^{\prime}$ signature are divided into
two categories: backgrounds that produce direct lepton pairs (referred to as direct-lepton backgrounds) and jet
backgrounds where one or both of the candidate leptons is from a misidentified jet.
Data events with an $\ell \ell^{\prime}$ invariant mass below 200~GeV constitute a control region to verify the background estimates,
and events with masses above 200~GeV comprise the signal search region.

The dominant direct-lepton backgrounds are $t\bar{t}$,
$Z/\gamma^* \rightarrow \ell \ell$, diboson ($WW$, $ZZ$, and $WZ$), and single top quark ($Wt$).
Since these processes are well understood and modelled, their contributions
are estimated using Monte Carlo samples generated
at $\sqrt{s}=7$ TeV and processed with the full \mbox{ATLAS} {\sc geant4}~\cite{geant}
simulation and reconstruction.
The event generators used are
{\sc pythia} 6.421~\cite{pythia} ($W$ and $Z/\gamma^*$), {\sc powheg} 1.0~\cite{powheg} ($t\bar{t}$),
{\sc madgraph 4}~\cite{madgraph} ($W/Z+\gamma$), {\sc mc@nlo} 3.4~\cite{mcnlo} (single top quark)
and {\sc herwig 6.510}~\cite{herwig} ($WW$, $WZ$ and $ZZ$). 
The parton distribution functions are CTEQ6L1~\cite{CTEQ} for $W$ and $Z$ production
and CT10~\cite{CT10} for $t\bar{t}$, single-top quark, and diboson production.
The Monte Carlo samples are normalized to cross sections with higher-order corrections applied. 
The cross section is calculated to next-to-next-to-leading order for $W$ and $Z/\gamma^*$~\cite{FEWZ},
next-to-leading order plus next-to-next-to-leading log for $t\bar{t}$
~\cite{Bonciani:1998vc}, and next-to-leading order for $WW$, $WZ$ and $ZZ$~\cite{MCFM}. 
Single-top quark and $W/Z+\gamma$ cross sections are calculated with {\sc mc@nlo} and {\sc madgraph}, respectively.
The effects of QED radiation are generated with {\sc photos}~\cite{PHOTOS}.
Hadronic tau decays are simulated with {\sc tauola}~\cite{TAUOLA}. 
Studies of leptons in $Z/\gamma^*$, $W$, and $J/\psi$ events~\cite{LeptonStudies} have shown that the lepton
reconstruction and identification efficiencies, energy scale, and energy resolution need small adjustment
in the Monte Carlo simulation to describe the data properly. The appropriate corrections are applied
to the Monte Carlo samples to improve the modelling of the backgrounds.
The effect of additional $pp$ interactions per bunch crossing as a function of the instantaneous
luminosity is modelled by overlaying simulated minimum bias events with the same distribution
in number of events per bunch crossing as observed in the data.

The processes $W/Z+\gamma$, $W/Z$+jets, and multijet production
give rise to backgrounds from 
jets misidentified as leptons, electrons from photon conversions, and leptons from hadron
decays (including $b$- and $c$-hadron decays).
The dominant component of these backgrounds is from events with one prompt lepton 
and one jet misidentified as a lepton, with an additional contribution from events with two misidentified jets. 
These backgrounds are estimated using data. The
background component initiated by prompt photons is estimated from Monte Carlo samples.

The jet backgrounds, including semileptonic decays in bottom and charm jets, are greatly reduced
by the lepton isolation and high-\pt\ requirements but are still significant.
The dominant jet background is due to
$W$+jets production, whose contribution is estimated using data from a subsample selected with the
same criteria as signal events but with the additional requirement \met~$>$~30~GeV.
This subsample is enriched in $W$+jets events, whose contribution is about 60\%, while the
multijet background is reduced to about 3\% and the direct-lepton background to about 37\%.
The potential effect of the multijet contribution
is included in the systematic uncertainty.
The contribution from direct-lepton backgrounds in the subsample is determined
from Monte Carlo simulation and is subtracted to give the number of $W$+jets events.
This number is extrapolated to the number in the full data sample without the \met\ criterion using the $W$+jets Monte Carlo samples.
The shapes of the $W$+jets background in various kinematic variables,
including $m_{\ell \ell^{\prime}}$, are taken from $W$+jets Monte Carlo samples.

Studies of event samples dominated by multijet events show that the probability that
a jet is misidentified as a lepton is independent of its charge~\cite{H_tautau},
with a 10\% uncertainty.
A same-sign sample is selected using the same criteria as for the signal sample but with
the sign requirement reversed.
The multijet background in the opposite-sign sample is taken to be equal to its contribution in the
same-sign sample.
Direct-lepton backgrounds produce more opposite-sign
than same-sign events, so the same-sign sample is enriched in multijet background.
Contributions to the same-sign sample by the direct-lepton backgrounds are determined from Monte Carlo simulation.
The $W$+jets contamination of the same-sign sample is determined by selecting
only same-sign events with $\met > 30$~GeV and then extrapolating to the full same-sign sample using Monte Carlo simulation.
The direct-lepton background and $W$+jets contributions are subtracted from the
observed same-sign sample to give the expected distribution and normalisation of the
multijet background in the opposite-sign sample.

Table~\ref{tab:data_mc_cutflow} shows the number of events selected
in data and the estimated background contributions with their uncertainties. 
The largest backgrounds in the signal region ($m_{\ell\ell^{\prime}} > 200$~GeV)
are $W$+jets events, arising primarily from the leptonic
decay of the $W$ and the misidentification of a jet as a lepton, and $t\bar{t}$ events, arising primarily
from semileptonic decays of both the $t$ and $\bar{t}$.
For the $e\tau_\mathrm{had}$ mode, there is a significant contribution from multijet events where
two jets are misidentified as leptons.  There is also a significant contribution
to the $e\mu$ mode from $WW$ diboson production where one $W$ decays to an electron
and the other to a muon.
Blank entries indicate an insignificant contribution to the background.
The dominant sources of systematic uncertainty for the background predictions arise
from the statistical uncertainty on the $W$+jets and multijet background
determinations from data, a 10\% uncertainty on extrapolation from the subsample to the
full sample in the
calculation of the $W$+jets backgrounds,
theoretical uncertainties on the cross sections of
the direct-lepton background processes (5\% to 10\%),
and the integrated luminosity uncertainty (3.9\%).
Other systematic uncertainties from the lepton trigger (1\%), 
the product of reconstruction and identification
efficiencies (1\%, 2\%, and 5\% for $e$, $\mu$, and $\tau$, respectively), and
the energy/momentum scale and resolution (1\%, 1\%, and 3\% for $e$, $\mu$,
and $\tau$, respectively) are small and have been included.
There are small correlations between the background estimates (for example,
from the luminosity), which are included when setting limits.

The expected number of events in the control region
agrees well with the observed number of events for all three
signatures ($e\mu$, $e\tau_\mathrm{had}$, and $\mu\tau_\mathrm{had}$).

\newcommand{\pho}{\phantom{0}}
\begin{table*}
\begin{center}
\caption{\label{tab:data_mc_cutflow} Estimated SM backgrounds and observed event yield for each signal category
for the background ($m_{\ell \ell^{\prime}} < 200$~GeV) and signal ($m_{\ell \ell^{\prime}} > 200$~GeV) regions.}
\begin{tabular}{@{} c|rrr|rrr}
\hline
    & \multicolumn{3}{c}{$m_{\ell \ell^{\prime}} < 200$~GeV} & \multicolumn{3}{|c}{$m_{\ell \ell^{\prime}} > 200$~GeV}  \\
\hline
  Process & $N_{e \mu}$ & $N_{e \tau_\mathrm{had}}$ & $N_{\mu \tau_\mathrm{had}}$ & $N_{e \mu}$ & $N_{e \tau_\mathrm{had}}$ & $N_{\mu \tau_\mathrm{had}}$ \\ \hline
 $Z/\gamma^* \rightarrow \tau \tau$    & $1880\pm150$ & $4300\pm600$  & $5300\pm600$  & $8\pm1$    & $24\pm3$   & $28\pm4$    \\
 $Z/\gamma^* \rightarrow e e$          &              & $1050\pm80$   &               &            & $44\pm3$   &             \\
 $Z/\gamma^* \rightarrow \mu \mu$      &              &               & $3030\pm290$  &            &            & $29\pm3$    \\
 $t\bar{t}$                            & $760\pm110$  & $96\pm18$     & $94\pm14$     & $251\pm30$ & $90\pm15$  & $70\pm13$   \\
 Diboson                               & $260\pm27$   & $57\pm8$      & $60\pm7$      & $71\pm8$   & $26\pm3$   & $24\pm3$    \\
 Single top quark                      & $87\pm8$     & $11\pm2$      & $9\pm1$       & $39\pm4$   & $10\pm2$   & $8\pm1$     \\
 $W$+jets                              & $420\pm260$  & $3500\pm700$  & $3200\pm600$  & $90\pm40$  & $370\pm80$ & $470\pm110$ \\
 multijet                              & $37\pm13$    & $2200\pm700$  & $730\pm230$   & $6\pm2$    & $150\pm50$ & $24\pm18$    \\ \hline
  Total   \\
  background                           & $3440\pm300$ & $11200\pm900$ & $12400\pm800$ & $460\pm60$ & $720\pm80$ & $650\pm90$  \\
\hline
  Data                                 & 3345         & 11212         & 12285         & 498        & 795        & 699         \\
\hline
\end{tabular}  
\end{center}
\end{table*}

\section{Signal simulation}
The production of an RPV $\tilde{\nu}_{\tau}$
followed by a lepton-flavour-violating decay into $e\mu$, $e \tau$, or $\mu \tau$ is considered
in the interpretation of the data.
The $\tilde{\nu}_{\tau}$ may be produced by either $d\bar{d}$ or $s\bar{s}$ but not $u\bar{u}$ annihilation.
This search is performed assuming exclusively $d\bar{d}$ production,
since $s\bar{s}$ production is expected to be a factor of 10 to 60 lower than $d\bar{d}$ production for the same couplings for sneutrino masses from 500~GeV to 2000~GeV.

In RPV SUSY, the LFV terms of the effective Lagrangian are given by
$\textrm{L} = \frac{1}{2} \lambda_{ijk} L_i L_j e_k
+ \lambda^{\prime}_{ijk} L_i Q_j d_k$, where $L$ and $Q$ are the lepton and quark
SU(2) supermultiplets, $e$ and $d$ are the lepton and down-like quark
singlet supermultiplets, and $i,j,k = 1,2,3$ refer to fermion generation number.
The theory requires $\lambda_{ijk} = - \lambda_{jik}$.
The $\lambda^{\prime}$ terms include coupling of downlike quark-antiquark pairs to sneutrinos,
and the $\lambda$ terms include couplings of the sneutrino to distinct charged leptons.
For the interpretation of this measurement, the sneutrino is produced by $d\bar{d}$ annihilation to \snutau\
with coupling $\lambda^{\prime}_{311}$ and decays to $\ell \ell^{\prime}$ with
couplings $\lambda_{132}$, $\lambda_{133}$, and $\lambda_{233}$ for $e\mu$, $e\tau$,
and $\mu\tau$, respectively.

The signal cross sections are calculated to next-to-leading order~\cite{rpv_sneutrino}
using CTEQ6L1 parton distribution functions~\cite{CTEQ} and
depend on the \snutau{} mass~($\msnt$), $\lambda^{'}_{311}$
and $\lambda_{i3k}$, where $i \neq k$ are the final-state lepton generations.
The sneutrino is assumed to be a narrow resonance, and the measurement here is
sensitive to the production coupling $\lambda^{\prime}_{311}$ and
the branching ratio $\snutau \to \ell \ell^{\prime}$. 
Monte Carlo events with \snutau\ decaying into $e\mu$, $e \tau$, and $\mu \tau$
are generated with {\sc herwig 6.520}~\cite{herwig,herwigSig} with 
sneutrino masses ranging from 500~GeV to 2000~GeV.

From precision low-energy experiments~\cite{low_energy}, the best limit on $\lambda^{\prime}_{311}$
is $0.012\times(m_{\tilde{d}}/100\textrm{ GeV}) = 0.12$ for the current lower limit on $m_{\tilde{d}}$.
The limit on $\lambda_{i3k}$ is $0.05\times(m_{\tilde{e}_k}/100\textrm{ GeV})$, where $\tilde{e}_k$ is the
$k^{\mathrm{th}}$ generation slepton.
Couplings of $\lambda^{\prime}_{311}= 0.11$, $\lambda_{i3k} = 0.07$ and
$\lambda^{\prime}_{311}= 0.10$, $\lambda_{i3k} = 0.05$ are used as benchmarks in this letter.
These are consistent with current limits and benchmarks used
in previous searches~\cite{RPVTevatron,ATLAS_emu}.  For these couplings,
the expected width of the sneutrino is approximately 0.1\% of its mass.
For the range of couplings considered in this letter, the width is always
less than 5\% of the mass.  If the couplings are significantly larger
than our benchmarks, the use of perturbation theory is not valid.

\section{Results}
\begin{figure}[pth]
\begin{center}
  \includegraphics[width=0.48\textwidth]{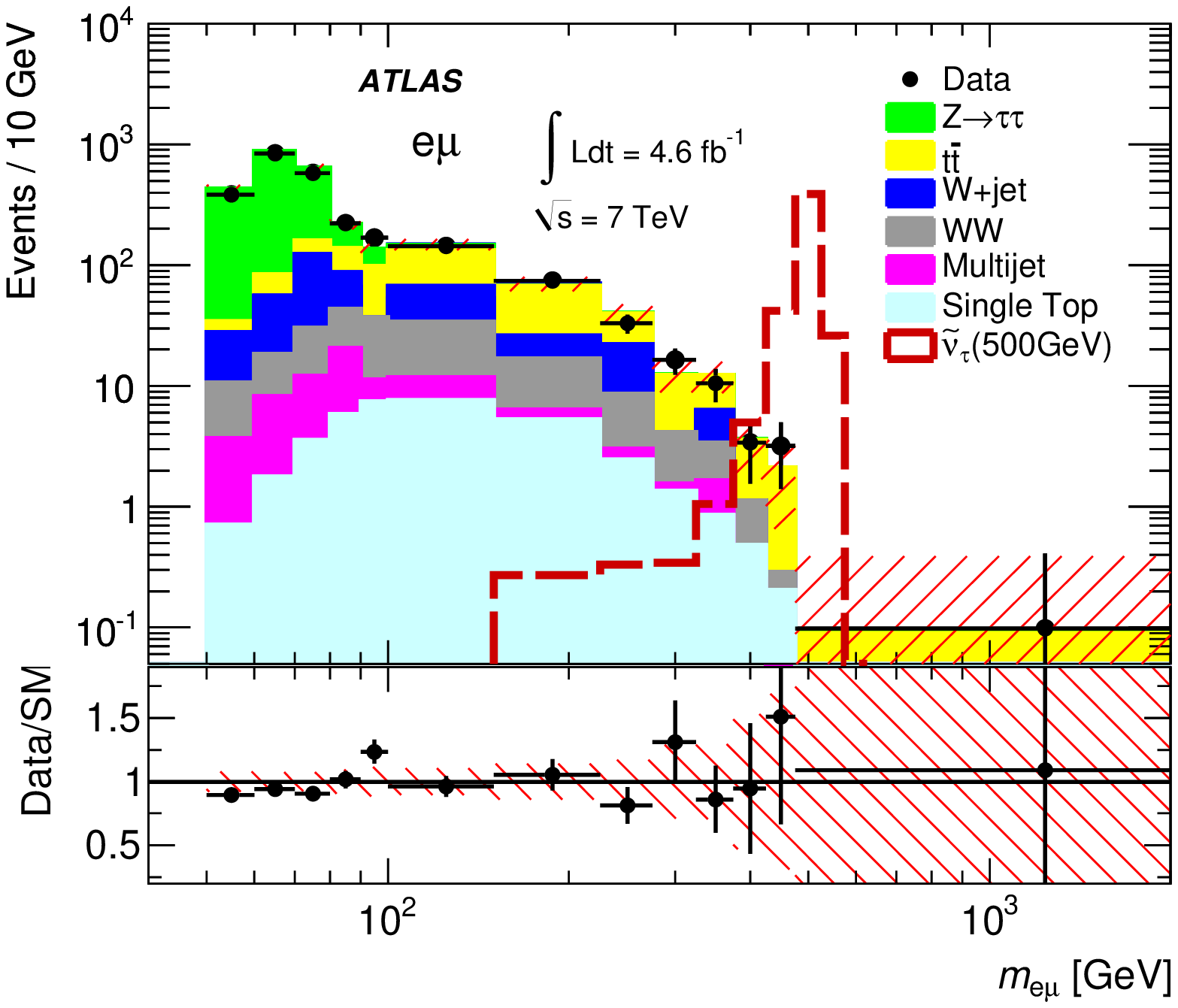} 
  \includegraphics[width=0.48\textwidth]{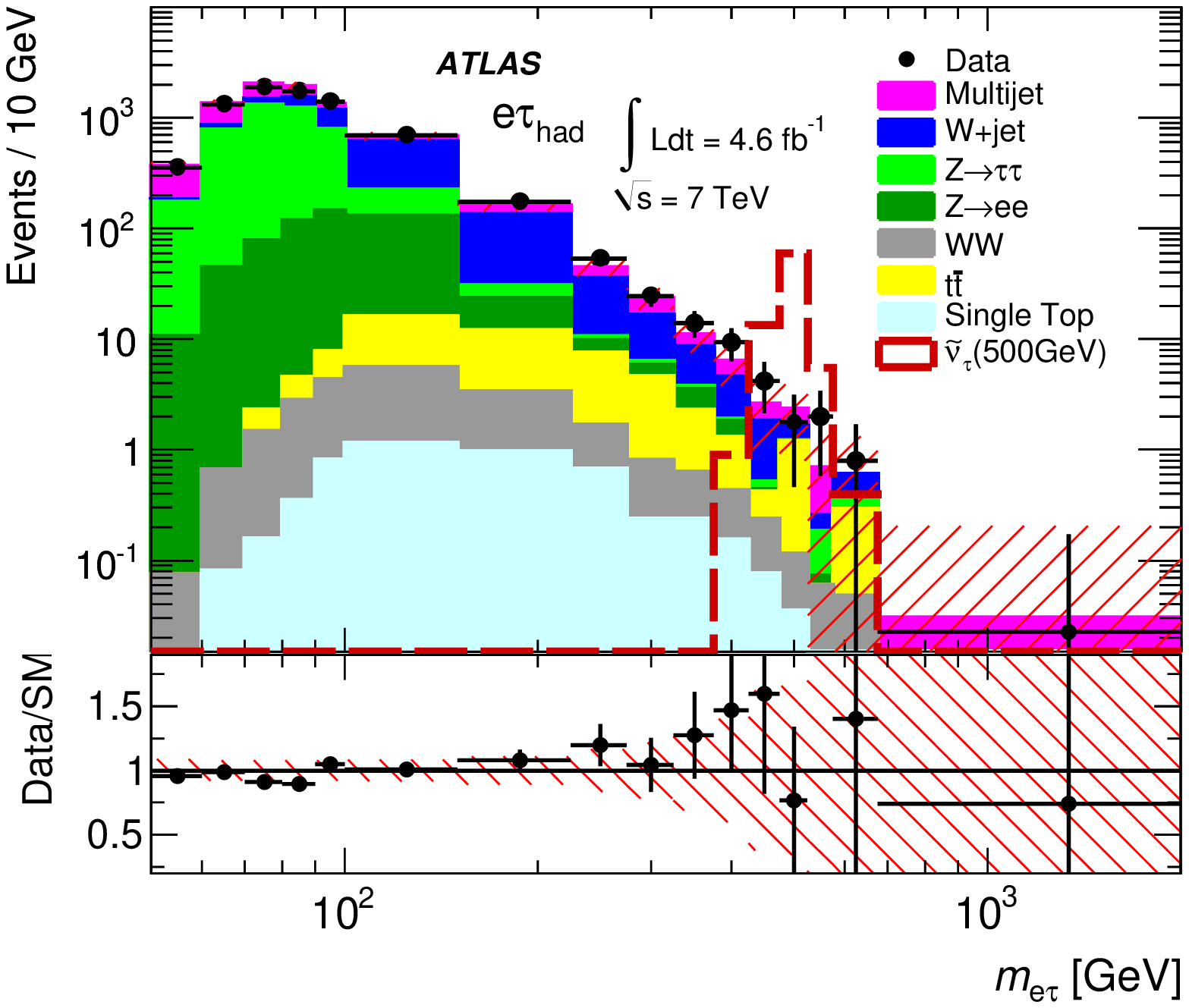} 
  \includegraphics[width=0.48\textwidth]{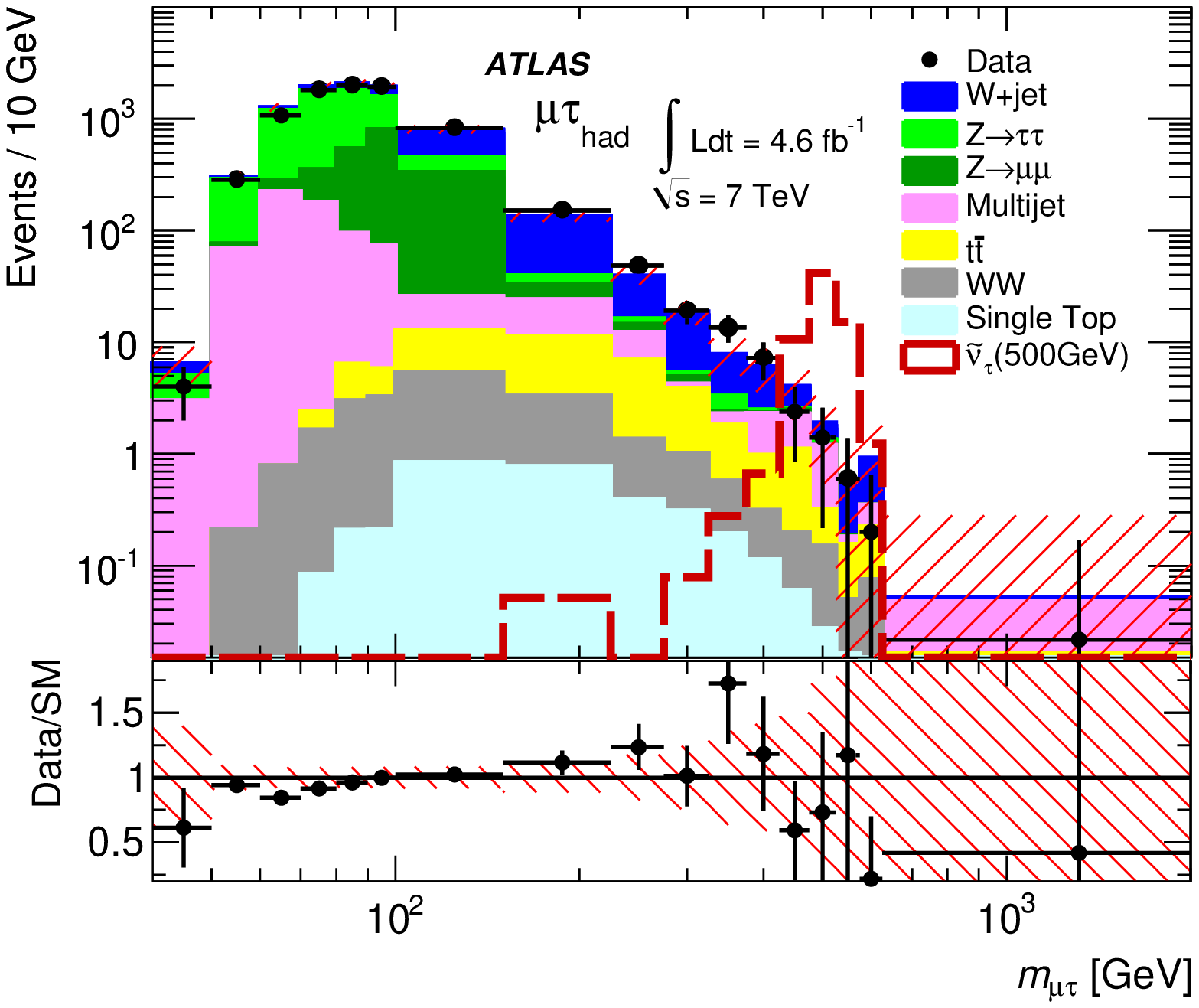}
\caption{Observed and predicted $\ell \ell^{\prime}$ invariant mass distributions
for $e\mu$ (top), $e\tau_\mathrm{had}$ (middle), and $\mu \tau_\mathrm{had}$ (bottom).
Signal simulations are shown
for $\msnt=500$~GeV ($\lambda'_{311}=0.11$, $\lambda_{i3k}=0.07$).
The region with $m_{\ell\ell^{\prime}}$~$<$~200~GeV is used to
verify the background estimation.
The lower plot for each decay mode shows the ratio of the data to the
SM backgrounds. The red hatching represents
the uncertainty on the total background in all plots.}
\label{fig:data_MC_Mass}
\end{center}
\end{figure}

The $\ell \ell^{\prime}$ invariant mass distributions in the signal region are presented
in figure~\ref{fig:data_MC_Mass} for data,
SM background contributions, and a \snutau~with
$\msnt=500$~GeV and with couplings
$\lambda^{\prime}_{311}= 0.11$ and $\lambda_{i3k} = 0.07$.

\begin{figure}[pth]
\begin{center}
   \includegraphics[width=0.48\textwidth]{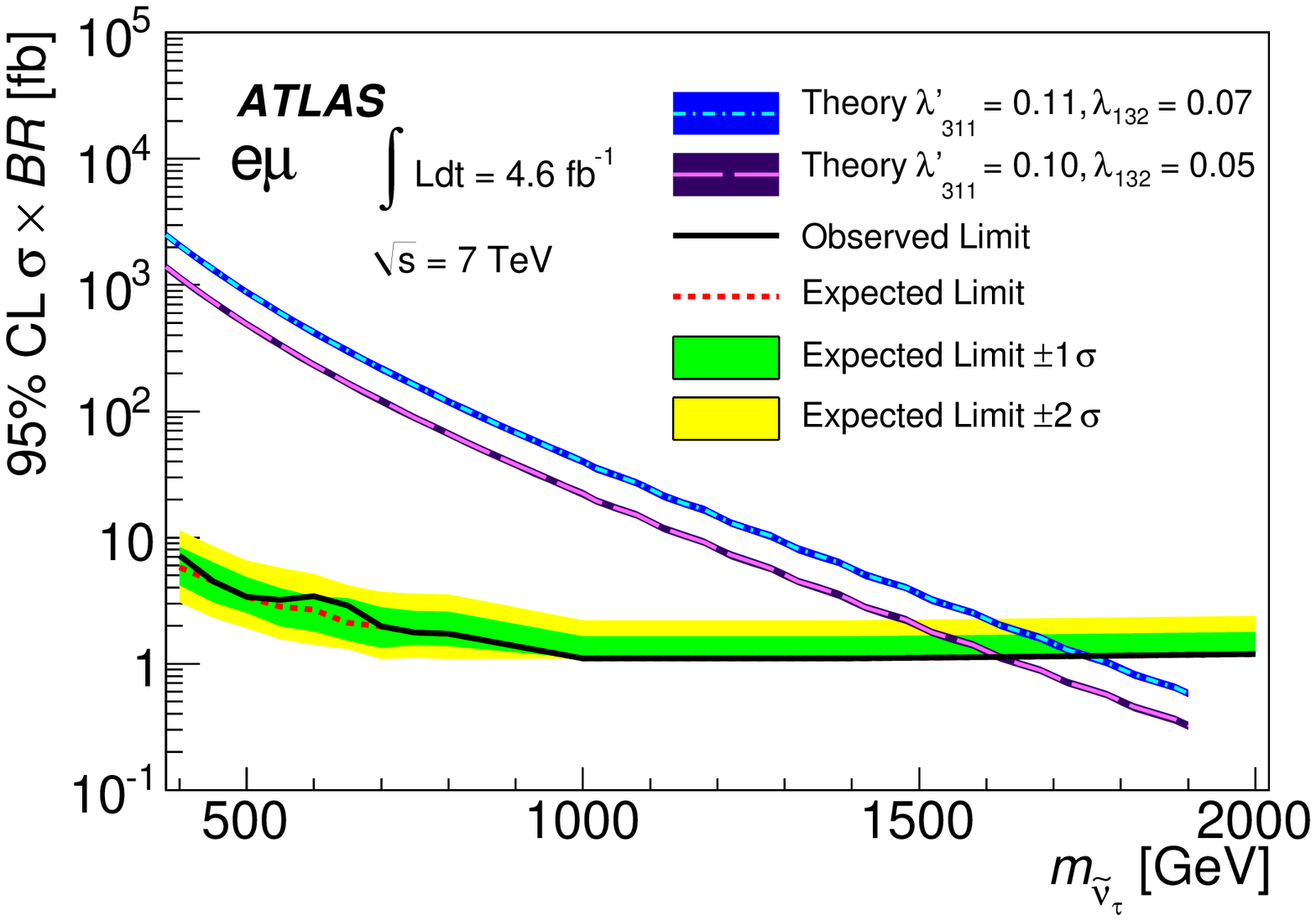}   \\
   \includegraphics[width=0.48\textwidth]{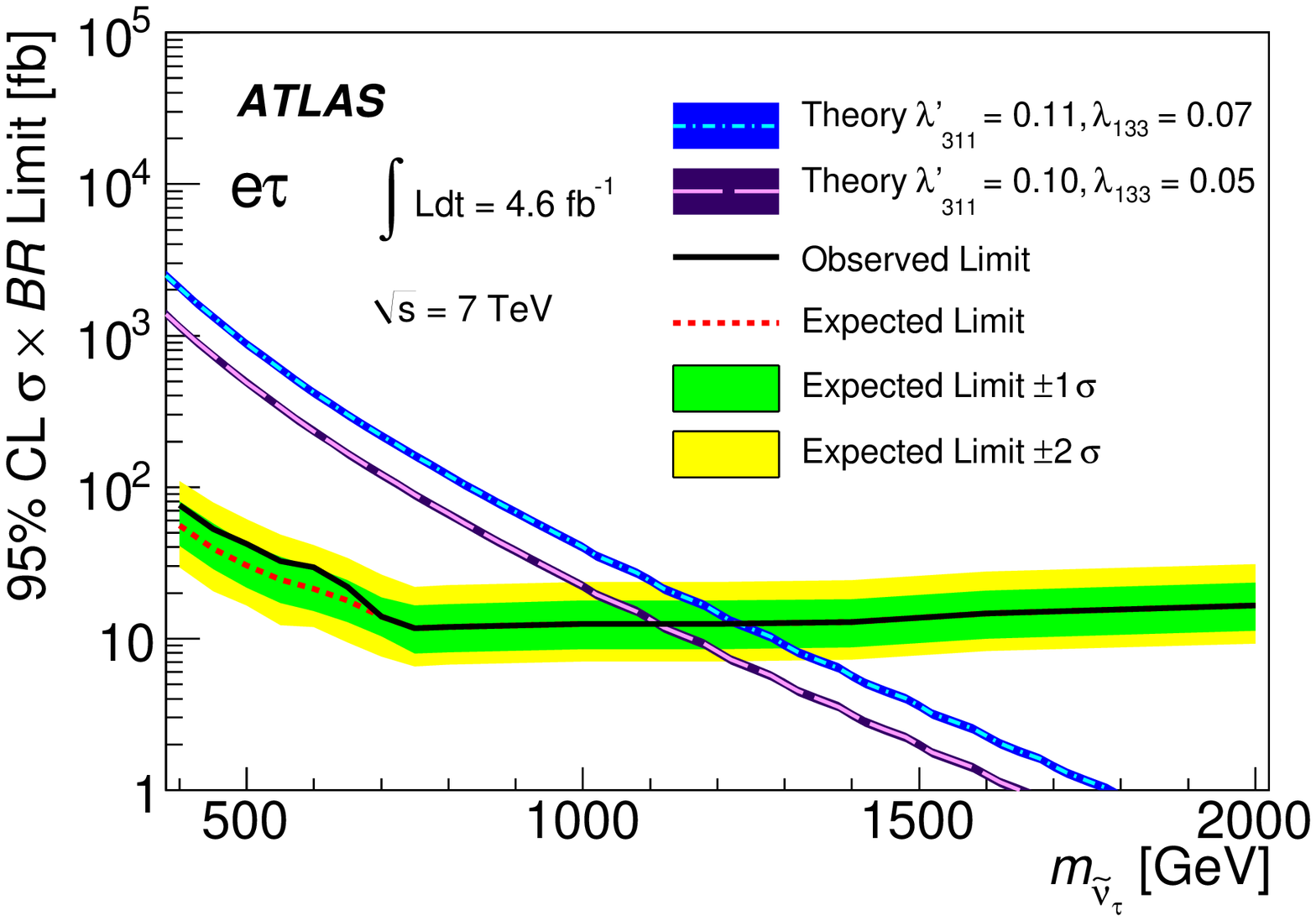}  \\
   \includegraphics[width=0.48\textwidth]{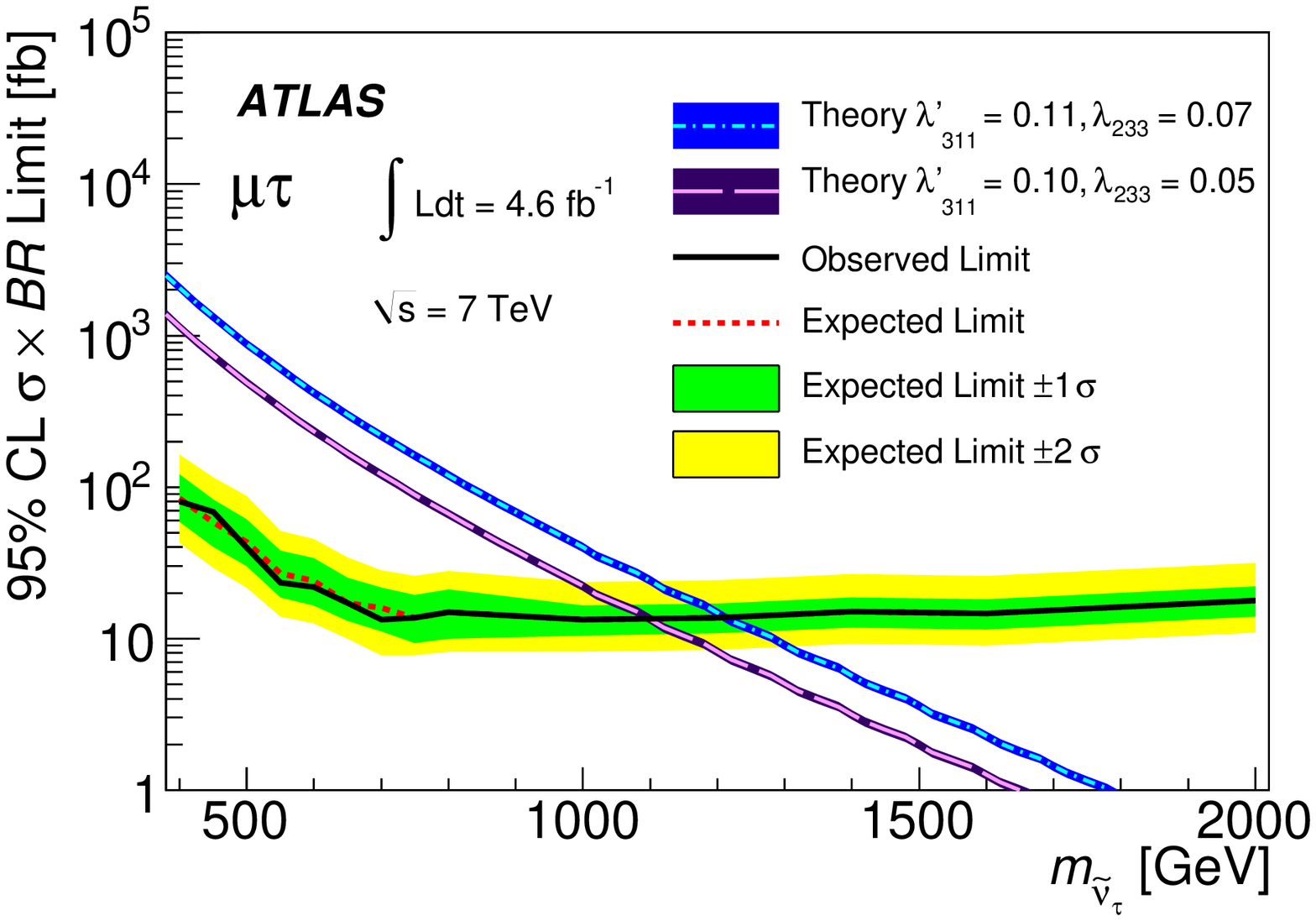}
\caption{The 95\% CL upper limit on the production cross section times branching
ratio as a function of sneutrino mass for $e \mu$ (top), $e \tau$ (middle), and
$\mu \tau$ (bottom) modes.
The red dotted curve is the expected limit,
the black solid curve is the observed limit,
and the yellow and green bands give
$\pm 1$ and $\pm 2$ standard deviations in the expected limit.
The expected theoretical curves for
$\lambda^{'}_{311}=0.11$, $\lambda_{i3k}=0.07$ (light blue dot--dash) and 
$\lambda^{'}_{311}=0.10$, $\lambda_{i3k}=0.05$ (light magenta dashed) are also plotted with their uncertainties. }
\label{fig:results_parameter_limits}
\end{center}
\end{figure}

The invariant mass spectra are examined for the presence of an RPV sneutrino.
No significant excess above the SM expectation is observed,
and limits are placed on the production cross section times branching ratio.
For each sneutrino mass, the search region is defined to be within $\pm$3 standard deviations
of the sneutrino mass, except
for $m_{\tilde{\nu}_{\tau}}$ above 800~GeV, where all events with $m_{\ell \ell^{\prime}} > 800$~GeV
are used.
The probability of observing a number of events as a function of the cross section times branching
ratio, efficiency, luminosity, and background expectation is constructed from a Poisson distribution.
The systematic uncertainties are included by convolution with Gaussian distributions.
The expected and observed 95\% confidence level (CL) upper limits on $\sigma(pp \rightarrow \tilde{\nu}_{\tau})
\times \textrm{BR}(\tilde{\nu}_{\tau} \rightarrow \ell \ell^{\prime})$
are calculated as a function of $\msnt$ using a Bayesian method~\cite{Bayesian} with a flat prior
for the signal cross section times branching ratio and integrating over the nuisance parameters.
Figure~\ref{fig:results_parameter_limits} shows the expected and observed limits as a function of $\msnt$,
together with the $\pm 1$ and $\pm 2$ standard deviation uncertainty bands. 
The expected exclusion limits are determined using
simulated pseudo-experiments containing only SM processes by
evaluating the 95\% CL upper limits for each pseudo-experiment at each 
value of $\msnt$, including systematic uncertainties.
The expected limit is calculated as the median of the distribution of limits.
The ensemble of limits is also used to find the $1\sigma$ and $2\sigma$ envelopes
of the expected limits as a function of $\msnt$. 
For a sneutrino mass of 500 (2000)~GeV, the observed limits on the production cross section
times branching ratio are 3.2~(1.4)~fb,
42~(17)~fb, and 40~(18)~fb for the $e\mu$, $e \tau$, and $\mu \tau$ modes, respectively.
The $e\tau$ and $\mu\tau$ limits are weaker because (1) the 1-track tau hadronic branching ratio is about 50\%,
(2) the tau reconstruction efficiency is lower due to criteria needed to reduce jet backgrounds,
and (3) the jet backgrounds are significantly larger than for the $e\mu$ mode.

In order to extract mass and coupling limits, it is assumed that only $d\bar{d}$ and $\ell\ell^{\prime}$
couple to the sneutrino.
The theoretical cross sections times branching ratios for $\lambda'_{311}=0.11$, $\lambda_{i3k}=0.07$
and $\lambda'_{311}=0.10$, $\lambda_{i3k}=0.05$ are also shown in figure~\ref{fig:results_parameter_limits}.
The branching ratio for each $\ell\ell^{\prime}$ mode depends on the couplings and is 21\%
for $\lambda'_{311}=0.11$, $\lambda_{i3k}=0.07$
and 14\% for $\lambda'_{311}=0.10$, $\lambda_{i3k}=0.05$.
The uncertainties on the theoretical
cross sections are evaluated by varying the factorisation and renormalisation scales (set equal to each other)
from $m_{\tilde{\nu}_{\tau}}/2$ to $2m_{\tilde{\nu}_{\tau}}$ and
varying the parton distribution functions.  These uncertainties are indicated as bands in
figure~\ref{fig:results_parameter_limits} and are small (only slightly larger than the width of the central line).
For couplings $\lambda'_{311}=0.10$, $\lambda_{i3k}=0.05$, the lower limits on the \snutau\ mass
are 1610~GeV, 1110~GeV, and 1100~GeV for $e \mu$, $e \tau$, and $\mu \tau$, respectively.
These lower limits are a factor of two to three higher than the best limits
from the Tevatron for the same couplings~\cite{RPVTevatron}.

\begin{figure}[pth]
\begin{center}
   \includegraphics[width=0.48\textwidth]{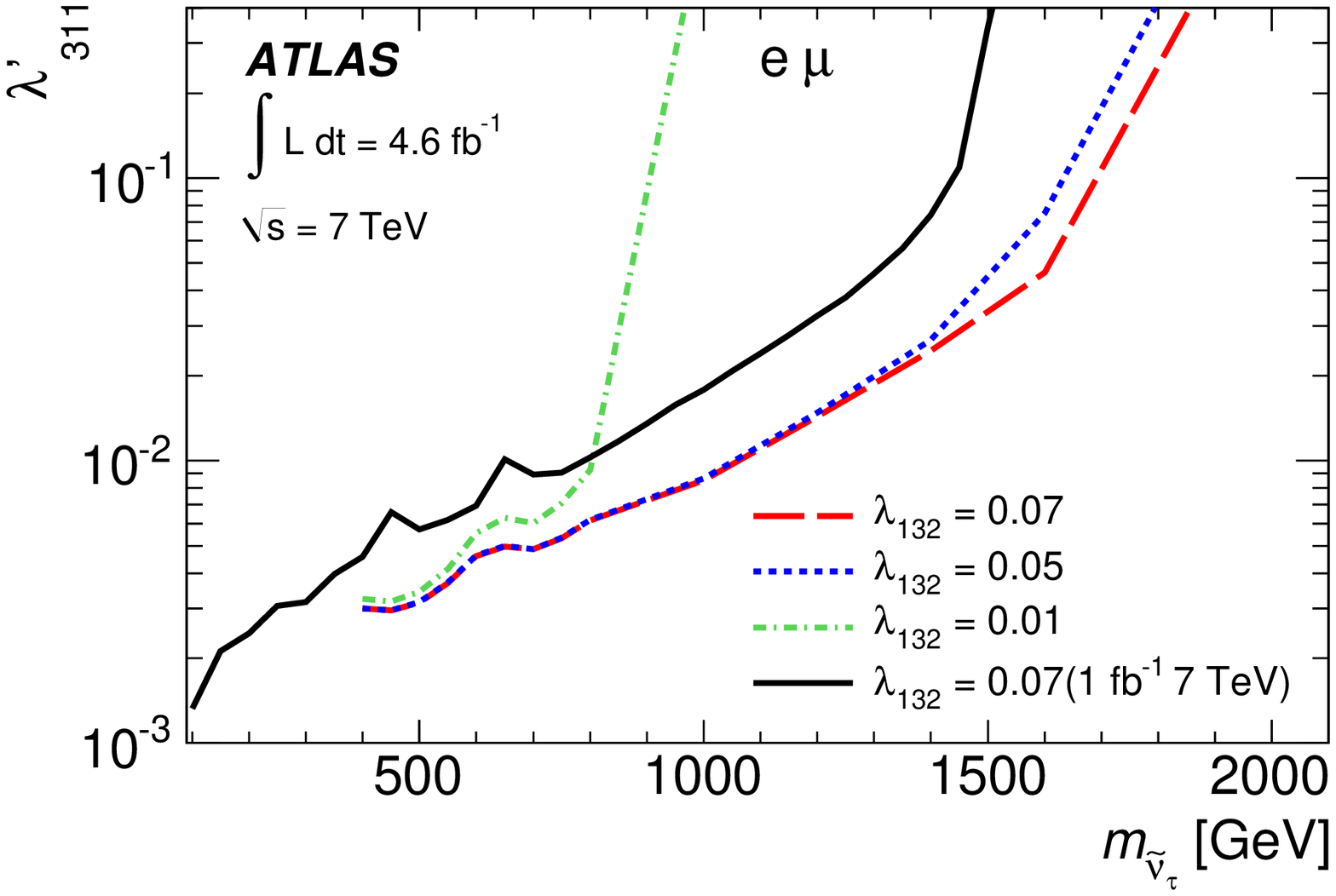}   \\
   \includegraphics[width=0.48\textwidth]{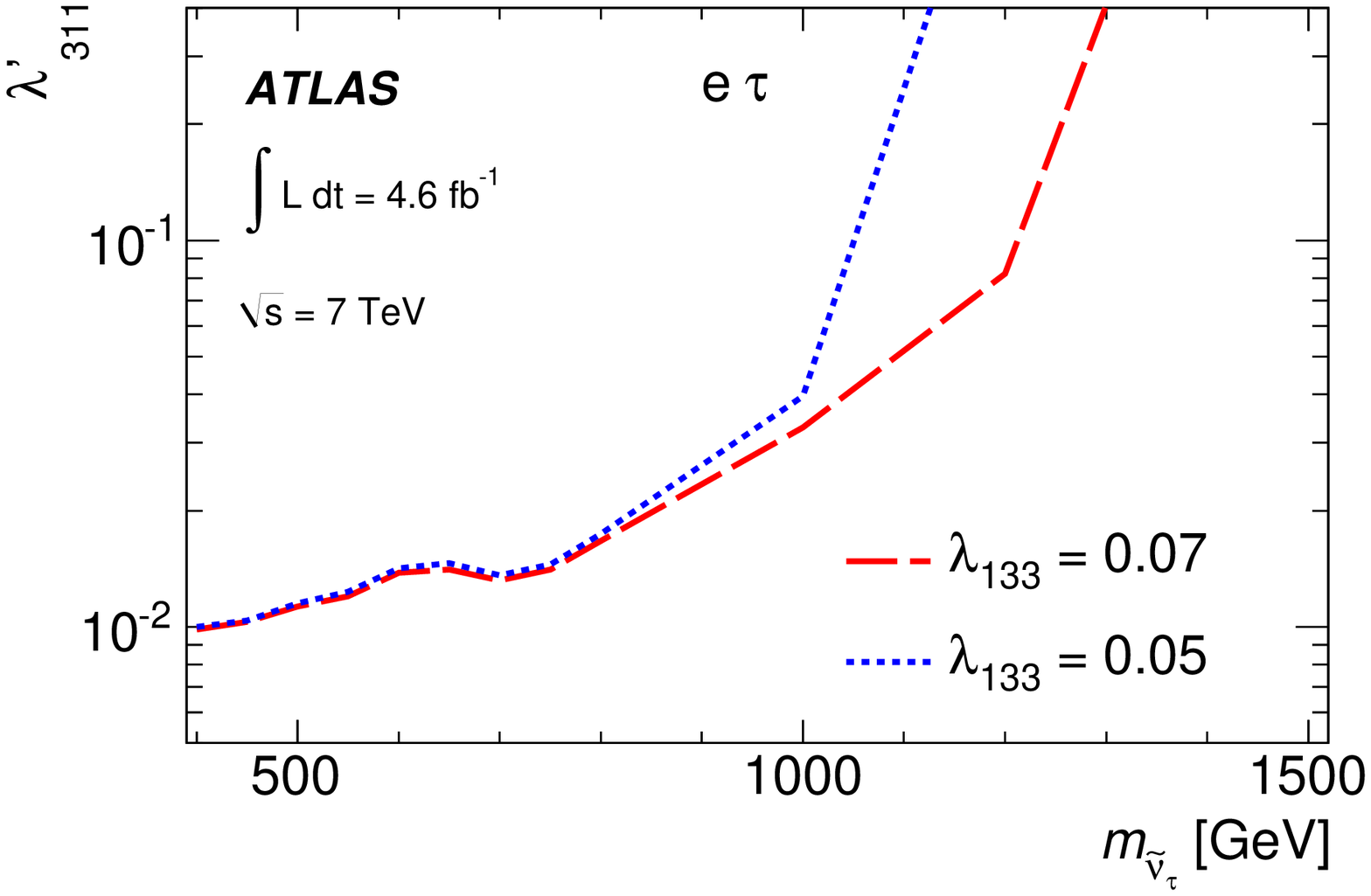}  \\
   \includegraphics[width=0.48\textwidth]{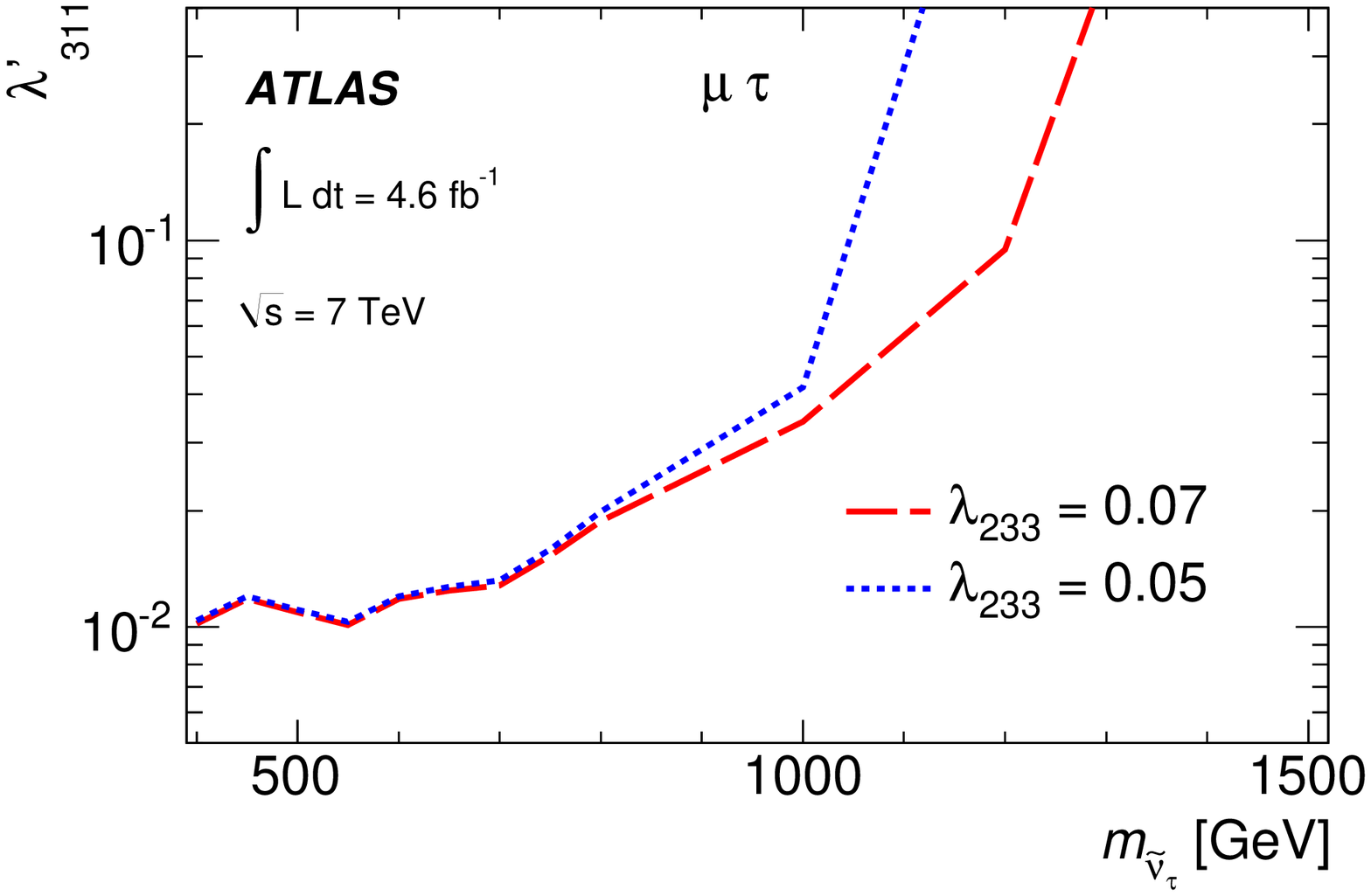}
\caption{The 95\% CL limits on $\lambda'_{311}$ as a function
of sneutrino mass for assumed values of $\lambda_{i3k}$ for the
$e \mu$ (top), $e \tau$ (middle), and
$\mu \tau$ (bottom) modes.
For the $e\mu$ mode, the black solid curve is the previous ATLAS result based
on 1~fb$^{-1}$ of data at 7 TeV.}
\label{fig:coupling_limits}
\end{center}
\end{figure}

The limits on the cross section times branching ratio are converted to limits on the couplings
under the assumption that there are no other significant couplings that contribute to the
decay of the \snutau.  In this case, the dependence of the cross section times branching ratio
on the couplings is
$|\lambda^{\prime}_{311}|^2 |\lambda_{i3k}|^2/(N_c|\lambda^{\prime}_{311}|^2 +2|\lambda_{i3k}|^2)$,
where the $|\lambda^{\prime}_{311}|^2$ in the numerator is from the production and the rest
is from the branching ratio.  The factor $N_c = 3$ is from color, and the 2 in the denominator comes
from accepting both charge states, that is, $\ell^+ \ell^{\prime -}$ and $\ell^- \ell^{\prime +}$.
~Figure~\ref{fig:coupling_limits} shows contours of the limit on $\lambda'_{311}$ as a function
of the sneutrino mass for various values of $\lambda_{i3k}$.  For each curve, the area
above the curve is excluded.  The previous limit from ATLAS for the $e\mu$ mode, 
based on 1~fb$^{-1}$ of 7 TeV data~\cite{ATLAS_emu}, is also shown.

\section{Summary}
A search has been performed for a heavy particle decaying to $e \mu$,
$e \tau_\mathrm{had}$, or $\mu \tau_\mathrm{had}$ final states using 4.6~fb$^{-1}$ of
$pp$ collision data at $\sqrt{s}=7$ TeV recorded by the \mbox{ATLAS} detector at the LHC.
The data are found to be consistent with SM predictions. 
Limits are placed on the cross section times branching ratio for an RPV SUSY sneutrino.
These results considerably extend previous constraints
from ATLAS~\cite{ATLAS_emu} and the Tevatron experiments~\cite{RPVTevatron}.

\section{Acknowledgements}
We thank CERN for the very successful operation of the LHC, as well as the
support staff from our institutions without whom ATLAS could not be
operated efficiently.

We acknowledge the support of ANPCyT, Argentina; YerPhI, Armenia; ARC,
Australia; BMWF, Austria; ANAS, Azerbaijan; SSTC, Belarus; CNPq and FAPESP,
Brazil; NSERC, NRC and CFI, Canada; CERN; CONICYT, Chile; CAS, MOST and
NSFC, China; COLCIENCIAS, Colombia; MSMT CR, MPO CR and VSC CR, Czech
Republic; DNRF, DNSRC and Lundbeck Foundation, Denmark; ARTEMIS, European
Union; IN2P3-CNRS, CEA-DSM/IRFU, France; GNAS, Georgia; BMBF, DFG, HGF, MPG
and AvH Foundation, Germany; GSRT, Greece; ISF, MINERVA, GIF, DIP and
Benoziyo Center, Israel; INFN, Italy; MEXT and JSPS, Japan; CNRST, Morocco;
FOM and NWO, Netherlands; RCN, Norway; MNiSW, Poland; GRICES and FCT,
Portugal; MERYS (MECTS), Romania; MES of Russia and ROSATOM, Russian
Federation; JINR; MSTD, Serbia; MSSR, Slovakia; ARRS and MVZT, Slovenia;
DST/NRF, South Africa; MICINN, Spain; SRC and Wallenberg Foundation,
Sweden; SER, SNSF and Cantons of Bern and Geneva, Switzerland; NSC, Taiwan;
TAEK, Turkey; STFC, the Royal Society and Leverhulme Trust, United Kingdom;
DOE and NSF, United States of America.

The crucial computing support from all WLCG partners is acknowledged
gratefully, in particular from CERN and the ATLAS Tier-1 facilities at
TRIUMF (Canada), NDGF (Denmark, Norway, Sweden), CC-IN2P3 (France),
KIT/GridKA (Germany), INFN-CNAF (Italy), NL-T1 (Netherlands), PIC (Spain),
ASGC (Taiwan), RAL (UK) and BNL (USA) and in the Tier-2 facilities
worldwide.

\clearpage
\onecolumn
\begin{flushleft}
{\Large The ATLAS Collaboration}

\bigskip

G.~Aad$^{\rm 48}$,
T.~Abajyan$^{\rm 21}$,
B.~Abbott$^{\rm 111}$,
J.~Abdallah$^{\rm 12}$,
S.~Abdel~Khalek$^{\rm 115}$,
A.A.~Abdelalim$^{\rm 49}$,
O.~Abdinov$^{\rm 11}$,
R.~Aben$^{\rm 105}$,
B.~Abi$^{\rm 112}$,
M.~Abolins$^{\rm 88}$
% ...
\bigskip
\\
% ...
$^{11}$ Institute of Physics, Azerbaijan Academy of Sciences, Baku, Azerbaijan\\
$^{12}$ Institut de F{\'\i}sica d'Altes Energies and Departament de F{\'\i}sica de la Universitat Aut{\`o}noma de Barcelona and ICREA, Barcelona, Spain\\
% ...
$^{21}$ Physikalisches Institut, University of Bonn, Bonn, Germany\\
% ...
$^{48}$ Fakult{\"a}t f{\"u}r Mathematik und Physik, Albert-Ludwigs-Universit{\"a}t, Freiburg, Germany\\
$^{49}$ Section de Physique, Universit{\'e} de Gen{\`e}ve, Geneva, Switzerland\\
% ...
$^{88}$ Department of Physics and Astronomy, Michigan State University, East Lansing MI, United States of America\\
% ...
$^{105}$ Nikhef National Institute for Subatomic Physics and University of Amsterdam, Amsterdam, Netherlands\\
% ...
$^{111}$ Homer L. Dodge Department of Physics and Astronomy, University of Oklahoma, Norman OK, United States of America\\
$^{112}$ Department of Physics, Oklahoma State University, Stillwater OK, United States of America\\
% ...
$^{115}$ LAL, Universit{\'e} Paris-Sud and CNRS/IN2P3, Orsay, France\\
% ...
\end{flushleft}


\end{document}